\documentclass[preprint,prb]{revtex4}

\usepackage{amsmath}
\newcommand{{\bff}}{\mathbf{f}}

\newcommand{\bfF}{\mathbf{F}}
\newcommand{\bfE}{\mathbf{E}}

\newcommand{\bfa}{\mathbf{a}}
\newcommand{{\bfg}}{\mathbf{g}}
\newcommand{{\bfv}}{\mathbf{v}}
\newcommand{{\bfA}}{\mathbf{A}}
\newcommand{{\bfV}}{\mathbf{V}}

\def\v#1{{\bf#1}}

\begin{document}

\title{Preacceleration without radiation: The non-existence of
preradiation phenomenon}
\author{Jos\'e A. Heras}
\email{heras@phys.lsu.edu}
\affiliation{Departamento de F\'\i sica, E. S. F. M., Instituto
Polit\'ecnico Nacional, M\'exico D. F. M\'exico and Department of Physics and Astronomy, Louisiana State University, Baton
Rouge, Louisiana 70803-4001, USA}
%\pacs{03.50.De, 03.50.Kk, 41.20.Cv, 41.20.Gz, 41.20.Jb}
\begin{abstract}
An unexpected prediction of classical electrodynamics is
that a charge can accelerate before a
force is applied. We would expect that a preaccelerated charge
would radiate so that there would be spontaneous preradiation, an acausal
phenomenon. We reexamine the subtle relation between the Larmor
formula for the power radiated by a point charge and the Abraham-Lorentz
equation and find that for well-behaved external forces acting for finite
times, the charge does not radiate in time intervals where there is
preacceleration. That is, for these forces preradiation
does not exist even though the charge is preaccelerated. The
radiative energy is emitted only in time intervals when the external force
acts on the charge. 
\end{abstract}
%\pacs{03.50.De, 03.50.Kk, 41.20.Cv, 41.20.Gz, 41.20.Jb}

\maketitle

\section{Introduction}
A universally accepted idea is that an accelerated charge radiates. However,
Abbott and Griffiths\cite{1} have found some extended configurations
with sinusoidal currents that do not radiate for certain special
frequencies for which the external fields are exactly zero. They refer to other work in which similar configurations involving acceleration without radiation has been discussed.\cite{1} The fact that acceleration
without radiation occurs when the external fields are exactly zero cannot be considered a fortuitous result attributable to extended configurations. We would expect that
a point charge and an extended charge would radiate only when an
external force acts on them. To understand why acceleration
without radiation is possible we should also consider the radiation reaction
force because the emission of radiation is generally
accompanied by a radiation reaction, a recoil force attributable to the
fields acting back on the charge. 

For a point charge $e$ of mass $m$ acceleration without radiation when the external force vanishes is not
supported by the Larmor formula for the instantaneous
power radiated:
\begin{equation}
P=m\tau a^2,
\end{equation}
where $\tau=2e^2/(3mc^3)$. The Abraham-Lorentz (AL) equation of motion
includes the radiation reaction force: 
\begin{equation}
m\bfa =\bfF + m\tau\dot{\bfa},
\end{equation}
where $\bfF(t)$ is an external force and a dot means time differentiation.
Equation (2) states that the charge can accelerate even when $\bfF$ vanishes
and it should radiate because of Eq.~(1). 

It is generally believed that radiation without the
intervention of an external force occurs in the preacceleration effect.\cite{2,3,4} To avoid runaway solutions in Eq.~(2) we can assume that $\bfa(\infty)=0$,\cite{3} but Eq.~(2) predicts preacceleration: the charge accelerates before the force acts on it. The charge should then radiate because of Eq.~(1). Therefore, Eqs.~(1) and (2) predict the existence of preradiation: the charge radiates before the external force acts on it. 

Although it is difficult to understand how acausal preaccelerations can exist
in the context of a classical theory, it is even more challenging to
understand how spontaneous preradiation can occur. Radiation in classical
electrodynamics is expected to be an observable quantity. But how can radiation
be detected when we have still not acted on the charge? No
direct or indirect experimental evidence of acausal preradiation
has been reported. The existence of preradiation
could be questioned because we can have acceleration without
radiation when the external fields are zero.\cite{1}

In this paper we reexamine the subtle relation between Eqs.~(1) and (2) and find that for well-behaved external forces acting for finite times, the charge does not radiate in time intervals where there is
preacceleration. That is, for these forces preradiation
does not exist even though the charge is preaccelerated. The
radiative energy is emitted only in time intervals when the external force
acts on the charge. 

\section{THE PRERADIATION PHENOMENON}
We review examples of preradiation for the following external forces: 
(A) Dirac delta force; (B) well-behaved force acting on a finite time; (C) periodical force acting on entire cycles and (D) constant force acting on a finite time.
Examples (A) and (D) are problems of Griffith's textbook.\cite{4,5,6,7}  Examples (C) and (D) are special cases of the example (B).

\subsection{The Dirac delta force}

Preradiation can be illustrated by an example due to
Dirac\cite{2} in which a charge is disturbed by a momentary pulse
represented by a delta function acting only at $t=0$:\cite{5}
\begin{equation}
F(t)= k \delta(t),
\end{equation}
where $k>0$. The nonrunaway solution of Eq.~(2) for this force is given by:\cite{5} 
\begin{equation}
a(t)= 
\begin{cases}
k e^{t/\tau}/(m\tau) & (t<0)\\
0. & (t>0)
\end{cases} 
\end{equation}
The total radiated energy in the interval $(t_1,t_2)$ is calculated with the formula
\begin{equation}
W_R(t_1,t_2)= m\tau\!\int_{t_1}^{t_2}a^2\,dt. 
\end{equation}
For the force in Eq.~(3) the total radiated energy is given by:\cite{5} 
\begin{equation}
W_R(-\infty,\infty)=W_R(-\infty,0)=\frac{k^2}{2m}, 
\end{equation}
which is radiated only during the preacceleration interval 
$(-\infty,0)$. Equation (6) is an example of $pure$ preradiation.

\subsection{Well-behaved external force}

Preradiation seems to exist for a more realistic external 
force beginning at $t=0$ and lasting until $t=T$,
\begin{equation}
\bfF(t)= 
\begin{cases}
0 & (t\leq 0) \\
\bff(t), & (0\leq t\leq T)\\
0 & t\geq T
\end{cases}
\end{equation}
where $\bff(t)$ is a well-behaved function of time. We note that Eq.~(7)
represents a family of forces. By assuming the condition
$\bfa(\infty)=0$, we can obtain the nonrunaway solution of Eq.~(2) for the
force in Eq.~(7):
\begin{equation}
\bfa(t)=
\begin{cases}
[\bfg(T)-\bfg(0)]e^{t/\tau}, & (t \leq 0) \\ 
[\bfg(T)-\bfg(t)]e^{t/\tau}, & (0\leq t\leq T) \\
0, & (t\geq T)
\end {cases}
\end{equation}
where the function $\bfg(t)$ is given by 
\begin{equation}
\bfg(t)=\frac{1}{m\tau}\!\int e^{-t/\tau}\bff(t)dt. 
\end{equation}
The total radiated energy $W_R(-\infty,\infty)$ is obtained by adding the
contributions in the intervals $(-\infty,0)$ and
$(0,T)$. Note that $W_R(T,\infty)=0$. From Eqs.~(5) and (8) we obtain 
\begin{equation}
W_R(-\infty,\infty)= \frac{m\tau^2}{2}[\bfg(T)-\bfg(0)]^2 +
m\tau\!\int_0^T [\bfg(T)-\bfg(t)]^2 e^{2t/\tau}\,dt.
\label{10}
\end{equation}
The first term in Eq.~\eqref{10} is the energy radiated in the interval $(-\infty,0)$ and the
second term is the energy radiated in the interval $(0,T)$. The external
force does not act in $(-\infty,0)$, but in this interval the charge is
preaccelerated (see Eq.~(8)) yielding the radiative energy appearing in the
first term of Eq.~(10), which represents another example of preradiation.

\subsection{Periodical external force}
As a special case of Eq.~(7) we consider the periodic force acting for
$n$ cycles: 
\begin{equation}
F(t)=
\begin{cases}
0, & (t\leq 0) \\
k\sin(\omega t), & (0\leq t\leq T)\\
0,& (t\geq T)
\end{cases}
\end{equation}
where $T=2n\pi/\omega$. In this case Eq.~(9)
takes the form
\begin{equation}
g(t)=-\frac{k
e^{-t/\tau}}{m(1+\omega^2\tau^2)}[\omega\tau\cos(\omega t)+\sin(\omega t)].
\end{equation}
The nonrunaway solution of Eq.~(2) when the external force is
specified by Eq.~(11) can be obtained from Eqs.~(8) and (12): 
\begin{equation}
a(t)= \begin{cases}
\alpha[ e^{t/\tau}- e^{(t-T)/\tau}], & (t\leq 0) \\
\alpha[\sin(\omega t)/\omega\tau+\cos(\omega t)- e^{(t-T)/\tau}], & (0\leq
t\leq T)\\
0, & (t\geq T)
\end{cases}
\end{equation}
where $\alpha=k\omega\tau/(m[1+\omega^2\tau^2])$. From Eqs.~(5) and (13) or
from Eqs.~(10) and (12) we can obtain the total radiated energy: 
\begin{equation}
W_R(-\infty,\infty)= \frac{m\tau^2 {\alpha}^2}{2}(1 -
e^{-T/\tau})^2+\frac{m\tau^2 \alpha^2}{2}\bigg(1+\frac{T}{\tau}+
\frac{T}{\tau^3\omega^2}- e^{-2T/\tau}\bigg),
\end{equation}
where the first term on the right-hand side is the preradiative energy
emitted during the interval $(-\infty,0)$ and the second term is the energy
radiated in the interval $(0,T)$. After a simple calculation, Eq.~(14)
becomes
\begin{equation}
W_R(-\infty,\infty)= \frac{m\tau^2 \alpha^2}{2}\bigg(2+\frac{T}{\tau}+
\frac{T}{\tau^3\omega^2}-2 e^{-T/\tau}\bigg).
\end{equation}

To verify that the energy is conserved in the interval $(-\infty,\infty)$
we first need to find the velocities associated with Eq.~(13). By assuming
$v(-\infty)=0$ we find 
\begin{equation}
v(t)= 
\begin{cases}
\alpha\tau( e^{t/\tau}-
e^{(t-T)/\tau}), & (t\leq 0) \\
\alpha [\omega\tau\sin(\omega t)-\cos(\omega t)-\omega^2\tau^2
e^{(t-T)/\tau}+\omega^2\tau^2+1]/\omega^2\tau, & (0\leq t\leq T)\\
0. & (t\geq T)
\end{cases}
\end{equation}
The work done by the periodic external force (11) is
\begin{subequations}
\begin{align}
W_{\rm ext}&= \!\int_{0}^{T}F(t)v(t)dt \\
&=\!\int_0^T\frac{k\sin(\omega t)\alpha}{\omega^2\tau}
(\omega\tau\sin(\omega t) -\cos(\omega t)-\omega^2\tau^2
e^{(t-T)/\tau}+\omega^2\tau^2+1)dt \\ &= \frac{m\tau^2
\alpha^2}{2}\bigg(2+\frac{T}{\tau}+ \frac{T}{\tau^3\omega^2}-2
e^{-T/\tau}\bigg).
\end{align}
\end{subequations}
Equation (17c) shows that the energy is conserved in the interval $(-\infty,\infty)$: the work done by the external force is equal to the change of kinetic energy (in this case it is zero because $v(\pm\infty)=0$) plus the radiated energy which is given by Eq.~(15).

\subsection{Constant external force}
As another special case of Eq.~(7) consider the constant force beginning at $t=0$ and lasting until $t=T$:\cite{6}
\begin{equation}
F= 
\begin{cases}
0, & (t<0) \\
k, & (0<t<T)\\
0. & (t>T)
\end{cases}
\end{equation}
This force is not continuous at $t=0$ and $t=T$ but its acceleration is continuous in these specific times.\cite{6} In this case Eq.~(9) becomes
\begin{equation}
g(t)=-\frac{k}{m} e^{-t/\tau}. 
\end{equation}
The corresponding nonrunaway solution of Eq.~(2) can be obtained from Eqs.~(8) and (19):\cite{6}
\begin{equation}
a(t)= 
\begin{cases}
k( e^{t/\tau}- e^{(t-T)/\tau})/m,
& (t\leq 0) \\
k (1- e^{(t-T)/\tau})/m, & (0\leq t\leq T)\\
0. & (t\geq T)
\end {cases}
\end{equation}
The energy radiated can be calculated with Eqs.~(5) and (20) or with
Eqs.~(10) and (19): 
\begin{equation}
W_R(-\infty,\infty)=\frac{k^2\tau^2}{2m}(1 - e^{-T/\tau})^2
+ \frac{k^2\tau^2}{2m}(-3+ 2T/\tau+ 4 e^{-T/\tau}- e^{-2T/\tau}).
\end{equation}
The first term on the right-hand side is the preradiative energy emitted during the interval $(-\infty,0)$ and the second term is the radiated energy during the interval $(0,T)$. After a simple calculation, Eq.~(21) becomes\cite{7}
\begin{equation}
W_R(-\infty,\infty)= \frac{k^2\tau}{m}(T-\tau+\tau e^{-T/\tau}).
\end{equation}
The integration of Eq.~(20) and the condition $v(-\infty)=0$ yields\cite{6} 
\begin{equation}
v(t)= 
\begin{cases}
k\tau( e^{t/\tau}- e^{(t-T)/\tau})/m,
& (t\leq 0) \\
k (t+\tau-\tau e^{(t-T)/\tau})/m, & (0\leq t\leq T)\\
kT/m, & (t\geq T)
\end {cases}
\end{equation}
The work done by the external force is then given by\cite{7}
\begin{subequations}
\begin{align}
W_{\rm ext}&= \!\int_{0}^{T}kv(t)dt\\
&=\!\int_0^T \frac{k^2}{m} (t+\tau-\tau 
e^{(t-T)/\tau})dt \\ &= \frac{k^2T^2}{2 m}+
\frac{k^2\tau}{m} (T-\tau+\tau e^{-T/\tau}). \label{24c}
\end{align}
\end{subequations}
The first term on the right-hand side of Eq.~\eqref{24c}
is the final kinetic
energy (the initial kinetic energy is assumed to be zero) and the second
term is the total radiated energy. Therefore Eq.~(24c) shows that the energy
is conserved, that is, the work done by the external force is equal to
the change of kinetic energy plus the total radiated energy.

\section{THE NON-EXISTENCE OF THE PRERADIATION PHENOMENON}
In Sec. II we have presented examples of acausal preradiative terms. However, acausal terms are physically unacceptable in purely classical considerations. Consistence of classical electrodynamics demands that acausal terms should not contribute to observable quantities.\cite{8} Therefore, the acausal preradiative terms should be natural and systematically eliminated from the total radiative energy. 

We note in particular that the second term in Eq.~(21) can be written as 
\begin{equation}
\frac{k^2\tau^2}{2m}(-3+ 2T/\tau+ 4 e^{-T/\tau}- e^{-2T/\tau})=\frac{k^2\tau}{m}(T-\tau+\tau e^{-T/\tau})
-\frac{k^2\tau^2}{2m}(1 - e^{-T/\tau})^2.
\end{equation}
The last term of Eq.~(25) exactly cancels the first (preradiative) term: $k^2\tau^2(1 - e^{-T/\tau})^2/(2m)$ of Eq.~(21) so that Eqs.~(21) and (25) imply Eq.~(22) for the total radiated energy. A similar cancellation of the preradiative term in Eq.~(14) occurs, that is, the second term of Eq.~(14) contains a part that exactly cancels its first (preradiative) term so that the total energy radiated is given by Eq.~(15). Equation (25)  seems to be the result of an integration by parts. Therefore, let us consider an integration by parts for the general case given by Eq.~(10). After integration by parts, we see that the second term in Eq.~(10): 
\begin{equation}
m\tau\!\int_0^T[\bfg(T)-\bfg(t)]^2 e^{2t/\tau}dt
=\tau\!\int_0^T[\bfg(T)-\bfg(t)] e^{t/\tau}\cdot\bff(t)dt
- \frac{m\tau^2}{2}[\bfg(T)-\bfg(0)]^2,
\end{equation}
contains a part that exactly cancels its first (preradiative) term, so that Eqs. (10) and (26) give the effective radiated energy 
\begin{equation}
W_R(-\infty,\infty)=\tau\!\int_0^T[\bfg(T)-\bfg(t)]
e^{t/\tau}\cdot\bff(t)\,dt.
\end{equation}
We can verify Eq.~(27) for the force in Eq.~(11). If we use Eqs.~(11) and
(12) in Eq.~(27), we obtain the result given in Eq.~(15):
\begin{subequations}
\begin{align}
W_R(-\infty,\infty)&=\tau\int_0^{T}\alpha\Big[\frac{\sin(\omega
t)}{\omega\tau}+\cos(\omega t)- e^{(t-T)/\tau}\Big] k\sin(\omega t)\,dt \\
&=\frac{m\tau^2
\alpha^2}{2}\Big[2+\frac{T}{\tau}+ \frac{T}{\tau^3\omega^2}-2
e^{-T/\tau}\Big].
\end{align}
\end{subequations}

Equations~(10) and (27) represent the same energy, but the trouble
with Eq.~(10) is that it explicitly exhibits a misleading time separation of
energy in the interval of preacceleration $(-\infty,0)$ and the
interval $(0,T)$, although Eq.~(27) shows that the energy is radiated
only during $(0,T)$ which is the interval when the force acts. We conclude
that there is no effective energy radiated during $(-\infty,0)$, which is
the interval when there is preacceleration. In other words: there does not
exist preradiation for the family of external forces
specified by Eq.~(7), even though the charge is preaccelerated. 

Rohrlich\cite{9} has emphasized that there are many equations
that have multiple solutions some of which are not realized in nature, and we
should distinguish between the existence of a mathematical solution and
its physical existence. For instance, the AL equation has two solutions:
$\bfa=0$ and $\bfa(t)=\bfa(0) e^{t/\tau}$, but only the former is physically possible. Advanced solutions
of Maxwell's equations violating the causality principle are another example
of solutions that are not physical. Analogously, acausal preaccelerations (of well-behaved external forces) 
exist mathematically (as an implication of the AL equation) but not physically because they do not originate radiation.

\section{Preradiation and the kinetic-radiative energy}
We note that the preradiative term in Eq.~(10) effectively depends on the value of the acceleration at $t=0$:
\begin{equation}
W_R(-\infty,0)=\frac{m\tau^2}{2}[\bfg(T)-\bfg(0)]^2=\frac{m\tau^2[a(0)]^2}{2}.
\end{equation}
It follows that from a formal point of view the elimination of preradiation implied by Eqs. (10) and (26) occurs at $t=0.$ To find how is the general dependence of the preradiative energies with respect the values of the acceleration,
 let us separate the interval of preacceleration $(-\infty,0)$ into the intervals $(-\infty,-t_0)$ and $(-t_0,0).$ After a simple calculation, we find $W_R(-\infty,-t_0)=m\tau^2[a(-t_0)]^2/2$ and $W_R(-t_0,0)=m\tau^2\left([a(0)]^2-[a(-t_0)]^2\right)/2$. The latter term depends on the value of the acceleration at the initial and final times of the interval $(-t_0,0).$  
But the time $-t_0$ is arbitrary and thus we have here a first lesson on preradiation: {\it the energy emitted during the interval of  preacceleration depends on the initial and final values of the acceleration in that interval.}

Equations (15) and (22) are examples of energy radiated by a charge
during the interval $(-\infty,\infty)$. To calculate these energies we have essentially combined Eqs.~(1) and (2), assumed the condition $\bfa(\infty)=0$ and integrated from $t=-\infty$ to $t=\infty$. The justification for the combination of Eqs.~(1) and (2) is that the emission of radiation is generally accompanied by a radiation reaction. The justification for the assumption $\bfa(\infty)=0$ is that it implies nonrunaway solutions. The justification for the integration over all time is that radiation is a process that occurs during a time interval rather than at one time. 

To generalize the method used to obtain Eqs. (15) and (22), we identify the acceleration in Eq.~(1) with that appearing in Eq.~(2) and obtain: 
\begin{equation}
P=\tau\bfa\cdot\bfF +\frac{d}{dt}\left(\frac{m\tau^2a^2}{2}\right).
\end{equation}
In analogy to the kinetic energy $T=mv^2/2$, we define the kinetic-radiative energy as 
\begin{equation}
T_{R}=\frac{m\tau^2a^2}{2}.
\end{equation}
Therefore
\begin{equation}
P=\tau\bfa\cdot\bfF +\frac{d  T_{R}}{dt}.
\end{equation}
The integration of Eq.~(32) over the interval $(t_1,t_2)$ gives the energy
radiated during that interval:
\begin{equation}
W_R(t_1,t_2)= \tau\!\int_{t_1}^{t_2}\bfa\cdot\bfF\, dt +\Delta T_{R}(t_1,t_2), 
\end{equation}
where $\Delta T_{R}(t_1,t_2)=T_R(t_2)-T_R(t_1).$ If the force vanishes in the interval $(t_1,t_2)$ then 
\begin{equation}
W_R(t_1,t_2)=\Delta T_{R}(t_1,t_2),
\end{equation}
according to which the radiated energy in $(t_1,t_2)$ is given by the change of the kinetic-radiative energy in that interval. In particular, preradiation occurs when the external force is zero and therefore the preradiative terms satisfy Eq. (34). We have here a second lesson on preradiation which makes more precise the first lesson: {\it the energy emitted in the interval of preacceleration $(t_1,t_2)$ is given by the change of the kinetic-radiative energy in that interval.}

A well-behaved force subject only to the condition that
it is switched on in the distant past and switched off in the distant future,
that is, $\bff(\pm\infty)=0$, is physically realizable. In solving Eq.~(2)
for the force in Eq.~(7) we imposed the condition\cite{3}
$\bfa(\infty)=0$ to obtain the nonrunaway solution (8), which 
implies the condition $\bfa(-\infty)=0$. In other words: the nonrunaway solutions of Eq. (2) for well-behaved forces satisfy the  boundary conditions  $\bfa(\pm\infty)=0,$\cite{10} which indicate that the charge begins and ends the motion as a free particle\cite{11} and imply $\Delta T_{R}(\pm\infty)=0$. We have here a third lesson on preradiation: {\it the nonrunaway solutions of the AL equation for well-behaved external forces  satisfy $\Delta T_{R}(\pm\infty)=0$ and therefore the associated preradiative energy is vanished. }

On the other hand, if the solution of Eq.~(2) in the interval $(t_1,t_2)$ is such that it vanishes at $t=t_1$ and
$t=t_2$ then $\Delta T_{R}(t_1,t_2)=0$ and Eq.~(31) reduces to the expression 
\begin{equation}
W_R(t_1,t_2)= \tau\!\int_{t_1}^{t_2}\bfa\cdot\bfF\, dt,
\end{equation}
which causally links the radiated energy with the external
force in such a way that {\it the energy is radiated only in the interval when
the force acts.} From Eqs. (5) and (35) we obtain 
\begin{equation}
m\tau\!\int_{t_1}^{t_2}a^2\,dt = \tau\!\int_{t_1}^{t_2}\bfa\cdot\bfF\, dt +\Delta T_{R}(t_1,t_2),
\end{equation}
which shows that the subtle difference between Eqs.~(5) and (35) is that the latter equation implicitly assumes $\Delta T_{R}(t_1,t_2)=0.$ 

Let us now apply Eq.~(35) to the family of forces specified by Eq.~(7). The associated nonrunaway solution satisfies $\Delta T_{R}(\pm\infty)=0$. If Eqs.~(7) and (8) are used in Eq.~(35)
with $t_1=-\infty$ and $t_2=\infty$, we obtain the result in Eq.~(27):
\begin{equation}
W_R(-\infty,\infty)=\tau\!\int_0^T [\bfg(T)-\bfg(t)]
e^{t/\tau}\cdot\bff(t)\,dt.
\end{equation}

\section{ENERGY CONSERVATION}

To verify that the energy is conserved in the time interval $(-\infty,\infty)$
if the external force is given by Eq.~(7), we need to find the
velocities associated with Eq.~(8). By assuming $\bfv(-\infty)=0$, we find 
\begin{equation}
\bfv(t)=
\begin{cases}
\tau[\bfg(T)-\bfg(0)] e^{t/\tau}, & (t\leq 0) \\
\bfV(t)-\bfV(0)+\tau [\bfg(T)-\bfg(t)] e^{t/\tau}, & (0\leq t\leq T)\\
\bfV(T)-\bfV(0),& (t\geq T)
\end {cases}
\end{equation}
where $\bfV(t)$ is the velocity of the charge when the radiation reaction is
not considered. The work done by the external force (7) is then given by
\begin{subequations}
\begin{align}
W_{\rm ext}&=\!\int_{-\infty}^{\infty}\bff(t)\cdot\bfv(t)dt
=\!\int_0^T\bff(t)\cdot\bfv(t)dt \\
&= \frac{m(\bfv(T))^2}{2} +\tau\!\int_0^T\bff(t)\cdot [\bfg(T)-\bfg(t)]
e^{t/\tau}dt. 
\end{align}
\end{subequations}
The first term on the right-hand side of Eq.~(39b) 
is the fina
kinetic energy (the initial is assumed to be zero) and the second term is
the total radiated energy given by Eq.~(37). Equation (39b) indicates 
that the energy is conserved in the interval $(-\infty,\infty)$, that is,
the work done by the external force equals the change of kinetic energy
plus the total radiated energy. It follows that {\it energy
conservation in Eq.~(39b) supports the conclusion that the energy is
radiated only during the interval when the force acts.} Applications of
Eq.~(39b) are given by Eqs.~(17c) and (24c).

\section{ DISCONTINUOUS FORCES}

We have successfully applied Eq.~(27) to the force (7) which is not
differentiable at $t=0$ and 
$t= T$. We can also apply Eq.~(27) to the force (18) which is not continuous
at $t=0$ and $t= T$. If we use Eqs.~(18) and (20) in Eq.~(27) and integrate
from $\varepsilon$ to $T-\varepsilon$, 
we have
\begin{equation}
W_R(-\infty,\infty)=
\!\int_{\varepsilon}^{T-\varepsilon}\frac{k^2\tau}{m}(1- e^{(t-T)/\tau})dt
=\frac{k^2\tau}{m}(-2\varepsilon + T+\tau e^{\varepsilon-T/\tau}-\tau
e^{-\varepsilon/\tau}).
\end{equation}
After taking the limit $\varepsilon\to 0$, we obtain the result given in
Eq.~(22).\cite{7} 

We should not apply directly Eq.~(27) to the delta function force in Eq.~(3)
because in this case the associated acceleration is not continuous at
$t=0$. In this case we can apply Eq.~(34) with $t_1=-\infty$ to
$t_2=-\varepsilon$. The result is
\begin{equation}
W_R(-\infty,-\varepsilon)=\frac{k^2}{2m} e^{-2\varepsilon/\tau},
\end{equation}
which represents apparent preradiation. However, we observe that $W_R$
in Eq.~(41) can be considered as an effective function of $\varepsilon$.
Therefore we can write $W_R=W_{\rm eff}(\varepsilon)=k^2
e^{-2\varepsilon/\tau} /(2m)$. In the limit 
$\varepsilon\to 0$, we obtain $W_{\it eff}(0)=k^2/(2m)$, that is, Eq.~(6).
Thus the energy is radiated at $t=0$, the only time at which the
delta function force acts and hence preradiation is also
eliminated in this case. Of course, Eq.~(27) and more generally Eq.~(33)
apply to continuously differentiable forces. 

\section{CONCLUDING REMARKS}

Preaccelerations are commonly ignored by appealing to the argument
that for electrons they occur in times so small (of
order $\tau=6.24\times 10^{-24}$\,s) that they would be practically
unobservable. Preaccelerations have also been connected with the point charge
model assumed in the derivation of the AL equation so that the
classical electrodynamics of a point charge seems to be a questionable
theory.\cite{12} The idea of abandoning the point charge model to avoid
preacceleration has been successfully developed by several
authors.\cite{12,13} Here we have demonstrated that for well-behaved
external forces acting over finite times, a preaccelerated point charge
does not have radiative effects, which is consistent with the idea that
preacceleration exists mathematically but not physically. Preradiation is eliminated because the nonrunaway solutions of the AL equation for well-behaved external forces satisfy $\Delta T_{R}(\pm\infty)=0$, where $T_R=m\tau^2a^2/2$ is the kinetic-radiative energy. The radiative energy is emitted only in time intervals when the external force
acts on the charge.

\begin{acknowledgments} 
The author thanks Professor R. F. O'Connell for useful discussions. He also thanks to the Department of Physics and Astronomy of the Louisiana State University for its hospitality.
\end{acknowledgments}


\begin{thebibliography}{99}
\bibitem{11}
The condition in the distant future $\bfa(\infty)=0$ in Eq.~(2)
fixes the condition in the distant past $\bfa(-\infty)=0$ if
well-behaved external forces, $\bff(\pm\infty)=0$, are considered. If instead,
we consider these same forces and assume $\bfa(-\infty)=0$, then the solution of Eq.~(2) in the distant future
becomes infinite, $\bfa(\infty)=\infty$. In this sense the physics of the AL equation 
is consistent with the idea that the future determines the
past. We note also that $\bff(\pm\infty)=0$ and
$\bfa(\pm\infty)=0$ are sufficient conditions to guarantee the vanishing of the radiation
reaction force $m\tau\dot\bfa$ at $t=\pm\infty$.
\end{thebibliography}
\end{document}